\documentclass[a4paper,11pt]{article}
\pdfoutput=1 

\usepackage{jheppub} 

\usepackage[T1]{fontenc} 

\usepackage{amssymb} 
\usepackage{amsmath} %
\usepackage{color}
\usepackage{booktabs} 
\usepackage{array} 
\usepackage{paralist} 
\usepackage{verbatim} 
\usepackage{subfig} 
\usepackage{placeins,tikz}
\usepackage{slashed}

\usepackage{hyperref}

\newcommand\encircle[1]{%
  \tikz[baseline=(X.base)] 
    \node (X) [draw, shape=circle, inner sep=0] {\strut #1};}

\newcommand{\be}{\begin{equation}}
\newcommand{\ee}{\end{equation}}

\newcommand{\N}{\mathcal{N}}

\date{} 

\begin{document}
\begin{titlepage}

\begin{center}
 {\LARGE\bfseries  
 Multi-centered ${\mathcal N}=2$ BPS black holes: 
 \\    \vskip 2mm
  a double copy description}
 \\[10mm]

\textbf{G.L.~Cardoso, S. Nagy and S.~Nampuri}

\vskip 6mm
{\em  Center for Mathematical Analysis, Geometry and Dynamical Systems,\\
  Department of Mathematics, 
  Instituto Superior T\'ecnico,\\ Universidade de Lisboa,
  Av. Rovisco Pais, 1049-001 Lisboa, Portugal}\\[10mm]

{\tt gcardoso@math.tecnico.ulisboa.pt}\;,\;\,{\tt  snagy@math.tecnico.ulisboa.pt}\;,\;\,{\tt
  nampuri@gmail.com}
\end{center}

\vskip .2in
\begin{center} {\bf ABSTRACT } \end{center}
\begin{quotation}\noindent 
We present the on-shell double copy dictionary for linearised ${\mathcal N}=2$ supergravity coupled to an arbitrary number of vector multiplets in four dimensions. Subsequently, we use it to construct a double copy description of multi-centered BPS black hole solutions in these theories in the 
weak-field approximation.

\end{quotation}
\vfill
\today
\end{titlepage}

\tableofcontents

\section{Introduction}\label{sect-introduction}

The search for a consistent description of gravitational degrees of freedom in terms of 
gauge field theoretic states has been one of the most enduringly challenging approaches to formulating a quantum theory of gravity. Hints of a gauge-gravity duality manifest themselves in string theory in two prominent forms. Firstly, they emerge in constructions of graviton amplitudes from gluon amplitudes \cite{Bianchi:2008pu,Bern:2008qj,Bern:2010ue,Chiodaroli:2014xia,Chiodaroli:2016jqw}, with and without supersymmetry. Secondly and most remarkably, they appear as a holographic duality relating the path integral of string theory in asymptotically AdS spacetimes with that of a lower dimensional 
CFT (AdS/CFT) \cite{Maldacena:1997re,Gubser:1998bc,Witten:1998qj}. The holographic approach has been effective in understanding non-perturbative gravitational states such as black holes in AdS spacetimes, while the amplitude calculations encode gauge-gravity relations in flat spacetime. In light of this,  very recently, the authors initiated a program in \cite{us} to capture information about the $4$d linearized ungauged ${\cal N}=2$ supergravity theory in terms of two distinct gauge field theories.\footnote{For a review of recent developments that identify symmetries of the supergravity
Lagrangian with those of field theory Lagrangians, see \cite{Borsten:2015pla}.}
Specifically, a double field dictionary was developed to describe on-shell configurations in a particular $4$d ${\cal N}=2$ supergravity theory described by the prepotential $F=- i X^0 X^1$, consisting of supergravity coupled to a single vector multiplet. We then successfully 
tested the dictionary on single-centered dyonic BPS black solutions in this theory in the weak-field approximation.\footnote{For a different approach, see \cite{Monteiro:2014cda,Luna:2015paa, Luna:2016due,Goldberger:2016iau}.} In this note, we take the inevitable next step in this program and generalize the results of \cite{us} by developing a double copy dictionary for on-shell configurations in  $4$d ${\cal N}=2$ supergravity theories coupled to an arbitrary number $n_V$ of vector multiplets. We use this dictionary to write down the explicit double copy expression for general multi-centered dyonic BPS black holes in these theories, in the weak-field approximation.

This dictionary makes use of a convolution $\star$
\cite{Anastasiou:2014qba}, by means of which one expresses 
gravitational field configurations $\varphi_G$ on the supergravity side in terms of field theory configurations 
$\varphi$ and $\tilde \varphi$,
\be  \varphi_G = \varphi \star  {\tilde \varphi} \;.
\label{convgg}
\ee
The double copy construction proceeds by tensoring an $\mathcal{N}=2$ super Yang-Mills (SYM) multiplet with an $\N=0$ bosonic sector consisting of a gauge field together with $n_V-1$ real scalars. At the level of momentum states, it was shown in 
\cite{Anastasiou:2015vba} that these are the multiplets that are relevant for the double copy construction of
these $\N=2$ supergravity theories. This is displayed in \autoref{table_on_shell_states}, where we give the helicity eigenstates that result from the tensoring. 

The fields of the $\N=2$ SYM multiplet will generally transform in the adjoint of a non-Abelian group $G$, and we allow the fields in the $\N=0$ bosonic sector to transform in the adjoint of some (possibly different) group $\tilde{G}$. 
A so-called spectator field will have to be included in the dictionary, in order to absorb the non-Abelian indices and reproduce the correct number of on-shell degrees of freedom contained in $n_V$ vector multiplets coupled to ${\mathcal N}=2$ supergravity.
This spectator field, which we will denote by $\phi$, transforms in the bi-adjoint representation of $G \times {\tilde G}$. In the presence of this spectator field, \eqref{convgg} is replaced by 
\be  
\label{generic_dictionary}
\varphi_G = \varphi^{\alpha} \star \phi_{{\alpha} \tilde \alpha} \star {\tilde \varphi}^{\tilde \alpha} \;.
\ee
By appropriately choosing the non-Abelian global groups, one ensures that all the independent on-shell degrees of freedom of the supergravity side
are captured by field theory degrees of freedom. Let's illustrate this with a simple example. 
In \autoref{table_on_shell_states} we omitted the explicit dependence on the spectator field, for the sake of notational 
simplicity. However, 
from \autoref{table_on_shell_states}, we see that part of the dictionary for the supergravity scalar fields $z^a$ will be of the form
\be
z^a\propto \sigma^\alpha\star\phi_{\alpha\tilde{\alpha}}\star\tilde{\sigma}^{\tilde{\alpha}a} \;,
\label{zdcu}
\ee
where $z^a$ and $\sigma$ are complex, and $\tilde{\sigma}^a$ are real. If we set $G=\tilde{G}=U(1)$, such that the summation over $\alpha$ and $\tilde{\alpha}$ is removed, we see that the right hand side of \eqref{zdcu}
does not contain enough independent degrees of freedom to describe the complex supergravity scalars $z^a$.
However, if we allow $G$ and $\tilde{G}$ to be non-Abelian, and given sufficiently large dimensions for these groups, the $z^a$ will be described by independent functions. In the double copy procedure, we will always assume that we have picked $G$ and $\tilde{G}$ such that no artificial constraints are imposed between the degrees of freedom of the gravitational theory. 

We now pause to comment on the spectator field: it was shown in \cite{Cachazo:2013iea} that the double copy relation, given at the level of integrands for tree-level S-matrices, assumes the form
\be
\mathcal{I}_{(\text{Grav})}\times\mathcal{I}_{(\Sigma)}=\mathcal{I}_{(\text{YM})} \times\mathcal{I}_{(\widetilde{\text{YM}})}, 
\ee 
where $\Sigma$ is a scalar in the bi-adjoint represenation of $G\times\tilde{G}$, the non-Abelian gauge groups of the two YM theories. A comparison with \eqref{generic_dictionary} (which, as we showed, is the correct expression for the field dictionary from the perspective of degrees of freedom counting) suggests that the spectator field we use is 
related to the 
biadjoint scalar found in \cite{Cachazo:2013iea,Hodges:2011wm,Bern:1999bx} as 
\be
\phi_{\alpha\tilde{\alpha}}=[\Sigma^{-1}]_{\alpha\tilde{\alpha}} \:,
\ee  
where $\Sigma^{-1}$ is the convolution inverse of the scalar field $\Sigma$, defined by $\Sigma^{-1}\star\Sigma=\delta$.
 
 For the sake of notational simplicity, we will choose to omit the explicit dependence on the spectator
 field throughout the rest of the paper.

The convolution $\star$ that appears in \eqref{convgg} is defined in Cartesian coordinates, and is given by
\be
[f\star g](x)=\int d^4y f(y)g(x-y) \;.
\ee
It satisfies the following property, which we will use repeatedly when deriving the on-shell double copy dictionary
for linearised ${\mathcal N}=2$ supergravity with vector multiplets, 
\be
\partial_\mu(f\star g)=(\partial_\mu f)\star g=f\star(\partial_\mu g) \;.
\label{shift-der}
\ee

\begin{table}[h]
\small
\begin{center}
 $\begin{array}{c|c|c|c}
 &\begin{array}{c}  \tilde{A}^-\end{array}& \begin{array}{c} \tilde{A}^+\end{array}&\begin{array}{c} 
 {\tilde \sigma}^a\end{array}\\
\hline
&&\\
\begin{array}{c}{A}^-\end{array}
&\begin{array}{cccccc} g^- &\end{array} 
&\begin{array}{cccccc}  \varphi_0\end{array}
&\begin{array}{c}  A^{a -}\end{array}
\\

&&\\
\begin{array}{c} \lambda_i^- \end{array}
&\begin{array}{cccccc} \psi_i^{-}\end{array} 
&\begin{array}{cccccc} \chi_i^+\end{array}
&\begin{array}{c}\chi_i^{a-} \end{array}
\\

&&\\
\begin{array}{c} \sigma^+,\sigma^- \end{array}
&\begin{array}{cccccc} A_{0,1}^-\end{array} 
&\begin{array}{cccccc} A_{0,1}^+\end{array}
&\begin{array}{c} z^{a +},z^{a -} \end{array}
\\

&&\\
\begin{array}{c}\lambda_i^+ \end{array}
&\begin{array}{cccccc} \chi_i^-\end{array} 
&\begin{array}{cccccc} \psi_i^{+}\end{array}
&\begin{array}{c}\chi_i^{a+} \end{array}
\\

&&\\
\begin{array}{c}{A}^+\end{array}
&\begin{array}{cccccc}   \varphi_1\end{array}
&\begin{array}{cccccc} g^+ &\end{array} 
&\begin{array}{c}{A}^{a+}\end{array}
\\

\end{array}$
\caption{\tiny{On-shell $(\mathcal{N}=2)_{SYM}\times[(\mathcal{N}=0)_{YM}+(n_V-1) {\tilde \sigma}]
= (\mathcal{N}=2)_{sugra}+n_V(\mathcal{N}=2)_{vector}$}}
\label{table_on_shell_states}
\end{center}
\end{table} 
\noindent

We use the superconformal approach to ${\mathcal N}=2$ supergravity theories
\cite{deWit:1979dzm, Bergshoeff:1980is, deWit:1980lyi, deWit:1984wbb,deWit:1984rvr}.
In this approach, the complex scalar fields appearing on the supergravity side are denoted by $X^I$, with
$I =0, \dots, n_V$. The physical scalar fields are denoted by $z^A = X^A/X^0$, with $A= 1, \dots, n_V$.
For the purpose of the double copy dictionary, we further split the latter set into $z^A =(z^1, z^a)$, where
$a= 2, \dots, n_V$.

The double copy dictionary is a dictionary for fluctuations around a fixed background.  On the supergravity
side, we take the background to be given by flat spacetime, allowing for the presence of constant scalar fields
which we denote by $\langle X^I \rangle$. On the field theory side, the background is also 
taken to be flat spacetime.
We then derive the double copy dictionary by linearising the theories around these
backgrounds. 
To keep the local symmetries manifest, 
we work with field strengths on the gravity side, and we exhibit
the double copy dictionary for these quantities.

In the following, we begin 
by displaying the linearised on-shell supersymmetry transformation rules that we will use to generate
the double copy dictionary for all the fields involved.\footnote{We use the conventions of 
\cite{Freedman:2012zz}.} Then, we proceed to explain our double copy ansatz.
Finally, we use the linearised supersymmetry transformation laws to work out the double copy relations
for the supergravity fields. 
We verify
that the linearised supersymmetry transformations on the super Yang-Mills side reproduce the linearised supergravity
transformation rules.  We refer to \autoref{App:dderiva} for a detailed derivation of the double copy dictionary.
Our on-shell dictionary is summarized in \eqref{the_dictionary_unpack}.
Finally, we use this dictionary \eqref{the_dictionary_unpack} to obtain an explicit double copy
description of multi-centered
dyonic BPS black hole solutions in ${\mathcal N} =2$ supergravity coupled to an arbitrary number of vector multiplets.
We refer to \autoref{App:multiBPSbh} for a brief review of some of the features of these black hole solutions.

\section{On-shell double copy dictionary for linearised ${\mathcal N}=2$ supergravity with vector
 multiplets}

We follow \cite{us} and use the conventions given there. In particular, we refer to Appendix A of \cite{us} for a summary of the features of the 
superconformal approach to ${\mathcal N}=2$ supergravity theories that we use.  In deriving the on-shell double copy dictionary, 
we will repeatedly use the property \eqref{shift-der} as well as the linearised equations of motion
 for the fields involved.
We refer to Appendix C of \cite{us} for a summary of the linearised equations of motion.
These have to be supplemented by the linearised equations of motion for the fields $\tilde \sigma^a$
in \autoref{table_on_shell_states}, i.e. 
$\Box \tilde \sigma^a =0$.

\subsection{Linearised on-shell supersymmetry transformation laws}

We begin by summarizing the linearised on-shell supersymmetry transformation rules that we will use to generate
the double copy dictionary for all the fields involved. As stated above, we will omit the dependence on
adjoint indices associated with the non-Abelian global group $G \times \tilde G$, for simplicity. We refer to \cite{us}, where this dependence is taken into account.

On the field theory side, 
a rigid $\mathcal{N}=2$ vector multiplet transforms as follows under on-shell supersymmetry transformations,
\be 
\label{rigid-transf}
\begin{aligned}
\delta A_\mu &= \frac{1}{2}\varepsilon^{ij}\bar{\epsilon}_i\gamma_\mu\lambda_j
+h.c.     \;, \\
\delta\lambda_i&=\gamma^\mu\partial_\mu\sigma\epsilon_i +\frac{1}{4}\gamma^{\mu\nu}F^{-}_{\mu\nu}\varepsilon_{ij}\epsilon^j   \;, \\
\delta\sigma &=\frac{1}{2}\bar{\epsilon}^i\lambda_i \;.
\end{aligned}
\ee

On the gravity side, the supergravity model is encoded in the prepotential function $F(X)$, with 
the complex scalar fields $X^I$ subjected to the Einstein frame constraint
\be
N_{IJ}X^I\bar{X}^J=-1 \;,
\label{einsfra}
\ee
where
\begin{equation}
N_{IJ} = - i \left( F_{IJ} - \bar F_{IJ} \right) \;\;\;,\;\;\; F_{IJ} = \frac{\partial^2 F(X)}{\partial X^I \partial X^J} \;.
\end{equation}
We linearise the supergravity theory
around a flat spacetime background with metric $\eta_{\mu \nu}$ and constant scalar fields $\langle X^I \rangle $,
\begin{eqnarray}
g_{\mu \nu} &=& \eta_{\mu \nu} + h_{\mu \nu} \;, \nonumber\\
X^I &=& \langle X^I \rangle + \delta X^I \;.
\label{flatbackg}
\end{eqnarray}
For notational simplicity, we will denote the fluctuations $\delta X^I$ simply by
$X^I$ in the following. Then, the linearised on-shell Q-supersymmetry transformation rules are given by
(dropping pure gauge terms in the variation of the gravitini)
\be
\label{all_sugra_SUSY}
\begin{aligned}
\delta_Q h_{\mu\nu}&=\bar{\epsilon}^i\gamma_{(\mu}\psi_{\nu)i}+h.c.   \;, \\
\delta_Q \psi_\mu^i&=-\frac{1}{4}\gamma^{ab}\partial_{[a}h_{b]\mu}^-\epsilon^i-\frac{1}{16}T_{\alpha\beta}^-\gamma^{\alpha\beta}\gamma_\mu\varepsilon^{ij}\epsilon_j  \;, \\
\delta_Q W_\mu^I&=\frac{1}{2}\varepsilon^{ij}\bar{\epsilon}_i\gamma_\mu\Omega^I_j  
+\varepsilon_{ij}\bar{\epsilon}^i\psi_\mu^j\langle \bar{X}^I\rangle +h.c. \;, \\
\delta_Q\Omega^{Ii}&=\gamma^\mu\partial_\mu\bar{X}^I\epsilon^i +\frac{1}{4}\gamma^{\mu\nu}
\mathcal{F}_{\mu\nu}^{I+}\varepsilon^{ij}\epsilon_j    \;, \\
\delta_Q X^I&=\frac{1}{2}\bar{\epsilon}^i\Omega_i^I \;,
\end{aligned} 
\ee   
with
\be
\label{comp_fields_lin}
\begin{aligned}
T_{ab}^{-} &= 2 \, \frac{\langle N_{IJ}  \, {\bar X}^J \rangle }{ \langle N_{KL} \,  {\bar X}^K {\bar X}^L\rangle } \, F_{ab}^{I-}  \;, \\
\mathcal{F}_{ab}^{I+}&= F_{ab}^{I+} - \tfrac12 \langle X^I\rangle \, T_{ab}^+  \;.
\end{aligned}
\ee
Note that the $U(1)$-connection, 
\be
a_{\mu} = - \frac12  \left( F_I \partial_{\mu} {\bar X}^I - {\bar X}^I \partial_{\mu} F_I + c.c. \right) \;,
\label{u1a}
\ee
vanishes at the linearised level, see \autoref{App:Kahlerconnec}.
The gaugini $\Omega^I_i$ are constrained by the linearised S-supersymmetry gauge fixing condition
\be
\langle {\bar X}^I \, N_{IJ} \rangle \, \Omega^J_i = 0 \;.
\label{s_super_lin}
\ee

\subsection{On-shell double copy dictionary \label{onshelldcdict}}

As in \cite{us}, we work with field strengths, such as 
$\psi^i_{\mu\nu} = 2 \partial_{[\mu}\psi^i_{\nu]}$, on the supergravity side. 
We follow \cite{us} and work in the Lorentz type gauge $\partial_{\mu} {\tilde A}^{\mu} =0$, for simplicity. In this gauge, we write
down the most general double copy ansatz for a linear combination of the supergravity fermions
that is compatible with \autoref{table_on_shell_states},
\be
\label{dict_ansatz} 
a\psi_{\mu\nu}^i+2b_I\gamma_{[\nu}\partial_{\mu]}\Omega^{Ii}
\equiv \varepsilon^{ij}\lambda_j \star\tilde{F}_{\mu\nu}  
+2c_a \gamma_{[\nu}\partial_{\mu]}\lambda^i\star\tilde{\sigma}^a \;,
\ee
where $a, b_I, c_a$ denote complex constants. Here we recall that the $\tilde \sigma^a$ denote $n_V -1$ real scalar
fields.

In the following, we will repeatedly use the property \eqref{shift-der} and the linearised equations of motion to extract information from \eqref{dict_ansatz}.
Contracting \eqref{dict_ansatz} with $\gamma^\mu$, using the equations of motion for $\psi_{\mu}^i$ and $\lambda_i$
as well as the property \eqref{shift-der},
we obtain
\be
\label{dict_gaugini}
2b_I\partial_\mu\Omega^{Ii}=\varepsilon^{ij}\gamma^\rho\lambda_j \star \partial_\mu\tilde{A}_\rho
+2c_a\partial_\mu\lambda^i\star\tilde{\sigma}^a \;.
\ee
Next,
we multiply \eqref{dict_gaugini} with $\gamma_\nu$ and anti-symmetrise to get
\be
\begin{aligned}
2b_I\gamma_{[\nu}\partial_{\mu]}\Omega^{Ii}
&=\varepsilon^{ij}\gamma_{[\nu}\gamma^\rho\lambda_j \star \partial_{\mu]}\tilde{A}_\rho
+2c_a \gamma_{[\nu}\partial_{\mu]}\lambda^i\star\tilde{\sigma}^a \\
&=\varepsilon^{ij}\lambda_j \star \tilde{F}_{\mu\nu}
-\varepsilon^{ij}\gamma^\rho\gamma_{[\nu}\lambda_j \star \partial_{\mu]}\tilde{A}_\rho
+2c_a \gamma_{[\nu}\partial_{\mu]}\lambda^i\star\tilde{\sigma}^a \;.
\end{aligned}
\ee
Now we substract the above from \eqref{dict_ansatz}  to obtain
\be 
\label{gravidc}
a\psi_{\mu\nu}^i=
\varepsilon^{ij}\gamma^\rho\gamma_{[\nu}\lambda_j \star \partial_{\mu]}\tilde{A}_\rho \;.
\ee
In \autoref{App:alterndcferm}  we show that the expression 
\eqref{gravidc} can be brought into the form
\be
\label{expressiongrav}
a\psi_{\mu\nu}^i =
\varepsilon^{ij}  \lambda_j \star  \tilde{F}^-_{\mu \nu}\;.
\ee
Then, 
inserting this into \eqref{dict_ansatz}, yields
\be 
2b_I\gamma_{[\nu}\partial_{\mu]}\Omega^{Ii}
= \varepsilon^{ij}\lambda_j \star\tilde{F}_{\mu\nu}^+  
+2c_a\gamma_{[\nu}\partial_{\mu]}\lambda^i\star\tilde{\sigma}^a \;,
\ee 
which upon contraction with $\gamma^\mu$ gives
\be
\label{expressiongg}
2b_I\partial_\mu\Omega^{Ii} =\varepsilon^{ij}\gamma^\rho\lambda_j \star \tilde{F}^+_{\mu \rho}
+2c_a\partial_\mu\lambda^i\star\tilde{\sigma}^a \;.
\ee
Expressions \eqref{expressiongrav} and \eqref{expressiongg} have to be
consistent with the linearised equations of motion for 
$\psi_{\mu}^i$, $\Omega^{Ii}$, ${\tilde A}_{\mu}$ and ${\tilde \sigma}^a$ . This is indeed the case, as can be easily checked, in a manner similar to the checks in \cite{us}.

Next, following \cite{us}, we derive the double copy relations for the other supergravity fields by applying  supersymmetry transformations to
the double copy relations  
\eqref{expressiongrav} and \eqref{expressiongg}. This will be discussed in \autoref{App:dderiva}.
We obtain the following dictionary,
\be
\label{the_big_dictionary}
\begin{aligned}
aR_{\mu\nu\alpha\beta}^-&=
-\frac{1}{2}\left[F_{\mu\nu} \star \tilde{F}_{\alpha\beta}^- +F_{\alpha\beta}^-\star \tilde{F}_{\mu\nu}
-4\eta_{[\alpha[\mu}\partial_{\nu]}\partial_{\beta]}^-A^\rho\star\tilde{A}_\rho\right] \\
a\psi_{\mu\nu}^i&=
\varepsilon^{ij}  \lambda_j \star  \tilde{F}^-_{\mu \nu}\\
aT_{\mu\nu}^-&=-4\sigma\star\tilde{F}_{\mu\nu}^-\\ 
b_I\mathcal{F}_{\mu\nu}^{I+}&=-\sigma\star\tilde{F}_{\mu\nu}^+
+c_a F_{\mu\nu}^+\star\tilde{\sigma}^a\\
b_I\partial_\mu\Omega^{Ii}&=
\varepsilon^{ij}\gamma^\rho\lambda_j \star \tilde{F}^+_{\mu \rho}
+2c_a\partial_\mu\lambda^i\star\tilde{\sigma}^a \\
b_I\partial_\mu\bar{X}^I&=\frac{1}{2}F_{\mu\rho}^-\star\tilde{A}^\rho+c_a \partial_\mu\bar{\sigma}\star\tilde{\sigma}^a 
\end{aligned}
\ee
In the expression for the Riemann tensor, the anti-selfdual part is taken over the indices $\alpha \beta$. The on-shell dictionary 
\eqref{the_big_dictionary} is invariant under local Abelian transformations $A \rightarrow A + d \alpha, {\tilde A} \rightarrow {\tilde A} + d {\tilde \alpha}$
by virtue of $\partial^{\mu} F_{\mu \nu} = 0$ and $\Box \tilde \alpha = 0$ (with the latter following from the Lorentz gauge condition $\partial^{\rho} \tilde A_{\rho} =0$).
In the absence of the fields $\tilde \sigma^a$, the dictionary \eqref{the_big_dictionary} reduces to the one given in \cite{us}.

Next, we need to extract the double copy relations for the individual fields 
$F_{\mu\nu}^I$, $\Omega^{Ii}$ and $X^I$ from \eqref{the_big_dictionary}. We begin by making the following 
general ansatz for the anti-sefldual part of $F_{\mu\nu}^I$,
\be
\label{ansatz_ind_F}
F_{\mu\nu}^{I-}=k^I\sigma\star\tilde{F}_{\mu\nu}^- +l^I\bar{\sigma}\star\tilde{F}_{\mu\nu}^-
+r^I_{\ a}F_{\mu\nu}^-\star\tilde{\sigma}^a \;,
\ee
with $k^I$, $l^I$ and $r^I_{\ a}$ complex constants. Inserting this ansatz into the expression for $T^-$ given in \eqref{comp_fields_lin}, and comparing
with the double copy relation for $T^-$ given in \eqref{the_big_dictionary}, we infer the constraints
\be 
\label{constraints_pqr_1}
\begin{aligned}
a\langle N_{IJ}\bar{X}^I \rangle k^J&=-2  \langle N_{KL} {\bar X}^K {\bar X}^L\rangle \;, \\
\langle N_{IJ}\bar{X}^I \rangle l^J&=0 \;, \\
\langle N_{IJ}\bar{X}^I \rangle r^J_{\ a}&=0 \;.
\end{aligned}
\ee
Next, we verify the compatibility with the double copy dictionary for $\mathcal{F}^+$. Using \eqref{comp_fields_lin} and \eqref{the_big_dictionary}, we have
\be
\label{curly_f1}
b_I\mathcal{F}_{\mu\nu}^{I+}=b_I F_{\mu\nu}^{I+} -\frac{1}{2}b_I\langle X^I\rangle T_{\mu\nu}^+  
=-\sigma\star\tilde{F}_{\mu\nu}^+ +c_a F_{\mu\nu}^+\star\tilde{\sigma}^a \;.
\ee
Inserting \eqref{ansatz_ind_F} and the dictionary for $T^+$ into this expression yields the constraints
\be
\label{constraints_pqr_2}
\begin{aligned}
b_I\left(\bar{k}^I+\frac{2}{\bar{a}}\langle X^I\rangle\right)&=0 \;, \\
b_I\bar{l}^I&=-1 \;, \\
b_I\bar{r}^I_{\ a}&=c_a \;.
\end{aligned}
\ee

Next, we read off the dictionary for the individual gaugini $\Omega^{Ii}$ from the variation of the field strengths  $F_{\mu\nu}^{I-}$. 
Using \eqref{sugrarelgg}, we have
\be
\label{sugra_field_ind}
\delta_Q F_{\mu\nu}^{I-} =\varepsilon^{ij}\bar{\epsilon}_i\gamma_{[\nu}\partial_{\mu]}\Omega^I_j
+\varepsilon_{ij}\bar{\epsilon}^i\psi_{\mu\nu}^j\langle \bar{X}^I\rangle \;,
\ee   
while from \eqref{ansatz_ind_F} we infer,
\be 
\delta_QF_{\mu\nu}^{I-}=k^I\delta_Q\sigma\star\tilde{F}_{\mu\nu}^- +l^I\delta_Q\bar{\sigma}\star\tilde{F}_{\mu\nu}^-
+r^I_{\ a}\delta_Q F_{\mu\nu}^-\star\tilde{\sigma}^a \;.
\label{dvvarFm}
\ee
Using the relations (see \autoref{App:dderiva})
\be 
\begin{aligned}
k^I\delta_Q\sigma\star\tilde{F}_{\mu\nu}^-
&=-\frac{a}{2}k^I\varepsilon_{ij}\bar{\epsilon}^i\psi_{\mu\nu}^j \;, \\
l^I\delta_Q\bar{\sigma}\star\tilde{F}_{\mu\nu}^-
&=-\frac{1}{2}l^I\varepsilon^{ij}\bar{\epsilon}_i\varepsilon_{jk}
\lambda^k\star\tilde{F}_{\mu\nu}^- \;, \\
r^I_{\ a}\delta_Q F_{\mu\nu}^-\star\tilde{\sigma}^a
&=r^I_{\ a}\varepsilon^{ij}\bar{\epsilon}_i\gamma_{[\nu}\partial_{\mu]}\lambda_j
\star\tilde{\sigma}^a \;,
\end{aligned}
\ee
 we get for \eqref{dvvarFm},
\be
\delta_QF_{\mu\nu}^{I-}=\varepsilon^{ij}\bar{\epsilon_i}
\left[-\frac{l^I}{2}\varepsilon_{jk}\lambda^k\star\tilde{F}_{\mu\nu}^- 
+r^I_{\ a}\gamma_{[\nu}\partial_{\mu]}\lambda_j\star\tilde{\sigma}^a\right] 
-\frac{a}{2}k^I\varepsilon_{ij}\bar{\epsilon}^i\psi_{\mu\nu}^j \;.
\ee
Comparing with \eqref{sugra_field_ind} determines the value of $k^I$,
\be
\label{constraint_3}
k^I=-2\frac{\langle\bar{X}^I\rangle}{a} \;,
\ee 
which satisfies the constraints
 \eqref{constraints_pqr_1} and \eqref{constraints_pqr_2},
 and also yields 
\be
\label{indgaugder}
\gamma_{[\nu}\partial_{\mu]}\Omega^I_j
=-\frac{l^I}{2}\varepsilon_{jk}\lambda^k\star\tilde{F}_{\mu\nu}^- 
+r^I_{\ a}\gamma_{[\nu}\partial_{\mu]}\lambda_j\star\tilde{\sigma}^a \;.
\ee 
Contracting \eqref{indgaugder} with  $\gamma^\mu$ 
and using
\be
\gamma^\mu \gamma_{[\nu}\partial_{\mu]}\Omega^I_j=-\partial_\nu \Omega^I_j \;,
\ee
and similarly for the second term on the right hand side of  \eqref{indgaugder}, we obtain the double copy relation
for the gaugini $\Omega^I_j $,
\be
\label{gaugini_individual_dict}
\partial_\mu\Omega^I_j=\frac{l^I}{2}\varepsilon_{jk}\gamma^\rho\lambda^k\star\tilde{F}_{\rho\mu}^- 
+r^I_{\ a}\partial_\mu\lambda_j\star\tilde{\sigma}^a \;.
\ee
Conversely. it can be checked that \eqref{gaugini_individual_dict} is consistent with \eqref{indgaugder}.
Moreover, \eqref{gaugini_individual_dict} 
has to be consistent with the dictionary entry in \eqref{the_big_dictionary}. Contracting 
\eqref{gaugini_individual_dict} with $b_I$ and comparing with
\eqref{the_big_dictionary} results in the conditions
$b_I\bar{l}^I=-1$ and $b_I\bar{r}^I_{\ a}=c_a$, which are precisely those in \eqref{constraints_pqr_2}.

Having determined the double copy expression for the gaugini $\Omega^I_i $, we now subject it to the
S-supersymmetry constraint \eqref{s_super_lin}, which results in the conditions 
$\langle{\bar X}^I N_{IJ} \rangle l^J=0$ and $\langle{\bar X}^I N_{IJ} \rangle r^J_{\ a}=0$, which are
precisely those in 
\eqref{constraints_pqr_1}.

By repeatedly making use of the equation of motion for $\lambda^k$ as well as
of $\partial_{\mu} {\tilde A}^{\mu } =0$, we derive the 
property 
\be
 \gamma^\rho\lambda^k\star\tilde{F}_{\rho\mu}^- =  \gamma^\rho\lambda^k\star\tilde{F}_{\rho\mu} \;,
 \ee
 which, when applied to \eqref{gaugini_individual_dict}, results in
 \be
\label{gaugini_individual_dict2}
\partial_\mu\Omega^I_i=-\frac{l^I}{2}\varepsilon_{ij}\gamma^\rho\lambda^j\star\partial_\mu\tilde{A}_\rho
+r^I_{\ a}\partial_\mu\lambda_i\star\tilde{\sigma}^a \;.
\ee
This will be used below to infer the double copy expression for the individual $X^I$, as follows.

We will use the notation $\delta_Q \varphi (\Sigma)$ to indicate that we are only considering terms in the
variation of a field $\varphi$ that are proportional to $\Sigma$. Consider then 
the Q-supersymmetry variation of $\Omega^{Ii}$.
Using \eqref{all_sugra_SUSY}, we have
\be 
\label{sugra_scalar_individual}
\delta_Q\partial_\mu\Omega^{Ii}(X)=\gamma^\rho\partial_\mu\partial_\rho\bar{X}^I\epsilon^i \;,
\ee
while using the double copy expression \eqref{gaugini_individual_dict2}, we have
\be
\begin{aligned}
\delta_Q\partial_\mu\Omega^{Ii}(X)&=
-\frac{\bar{l}^I}{2}\varepsilon^{ij}\gamma^\rho\delta_Q\lambda_j(F)\star\partial_\mu\tilde{A}_\rho
+\bar{r}^I_{\ a}\partial_\mu\delta_Q\lambda^i(\sigma)\star\tilde{\sigma}^a \\
&=\frac{\bar{l}^I}{8}\gamma_\rho\gamma^{\alpha\beta}F_{\alpha\beta}^-\star\partial_\mu\tilde{A}_\rho\epsilon^i
+\bar{r}^I_{\ a}\gamma^\rho\partial_\mu\partial_\rho\bar{\sigma}\star\tilde{\sigma}^a \epsilon^i \;.
\end{aligned} 
\ee
Using the relation
\be
\label{app_anti_self_contract}
\gamma_\rho\gamma^{\alpha\beta}F_{\alpha\beta}^- \, \epsilon^i =-4\gamma^\alpha F_{\alpha\rho}^- \, \epsilon^i \;,
\ee
we obtain
\be 
\begin{aligned}
\delta_Q\partial_\mu\Omega^{Ii}(X)
&=\gamma^\rho\partial_\mu\left(-\frac{\bar{l}^I}{2}F_{\rho\nu}^-\star\tilde{A}^\nu
+\bar{r}^I_{\ a}\partial_\rho\bar{\sigma}\star\tilde{\sigma}^a
\right)\epsilon^i \;.
\end{aligned}
\ee
Comparing with \eqref{sugra_scalar_individual}, we infer the double copy expression for $X^I$,
\be
\label{dict_scalar_individual}
\partial_\mu\bar{X}^I= -\frac{\bar{l}^I}{2}F_{\mu\rho}^-\star\tilde{A}^\rho
+\bar{r}^I_{\ a}\partial_\mu\bar{\sigma}\star\tilde{\sigma}^a \;.
\ee
This has to be consistent with the dictionary entry in \eqref{the_big_dictionary}. Contracting 
\eqref{dict_scalar_individual}  with $b_I$ and comparing with
\eqref{the_big_dictionary} results again in the conditions
$b_I\bar{l}^I=-1$ and $b_I\bar{r}^I_{\ a}=c_a$, which are precisely those in \eqref{constraints_pqr_2}.

Next, we verify the consistency of \eqref{dict_scalar_individual} with the Einstein constraint \eqref{einsfra}.
Differentiating \eqref{einsfra} once and linearising, we obtain
\be
\partial_\mu N_{IJ}\langle X^I\bar{X}^J\rangle 
+\langle N_{IJ}\bar{X}^J\rangle\partial_\mu X^I
+\langle N_{IJ}X^I\rangle\partial_\mu\bar{X}^J=0 \;,
\ee
which is satisfied by virtue of the special geometry relation $F_{IJK} X^K = 0$ and 
by virtue of the constraints \eqref{constraints_pqr_1}.

It is straightforward to check the consistency of the double copy expressions \eqref{ansatz_ind_F}, 
\eqref{gaugini_individual_dict} and  \eqref{dict_scalar_individual} with the equations of motion for
the various fields involved. In addition, one can show that both sides of these expressions transform identically
under supersymmetry. This is shown in \autoref{App:dderiva}.

We summarize the resulting on-shell double copy dictionary for all the supergravity fields,\footnote{We remind the reader that the convolution is taken over a bi-adjoint spectator scalar $\phi_{\alpha\tilde{\alpha}}$, as explained in \autoref{sect-introduction}. This is omitted in \eqref{the_dictionary_unpack} for the sake of clarity.} 
\be
\label{the_dictionary_unpack}
\boxed{
\begin{aligned}
aR_{\mu\nu\alpha\beta}^-&=
-\frac{1}{2}\left[F_{\mu\nu} \star \tilde{F}_{\alpha\beta}^- +F_{\alpha\beta}^-\star \tilde{F}_{\mu\nu}
-4\eta_{[\alpha[\mu}\partial_{\nu]}\partial_{\beta]}^-A^\rho\star\tilde{A}_\rho\right] \\
a\psi_{\mu\nu}^i&=
\varepsilon^{ij}  \lambda_j \star  \tilde{F}^-_{\mu \nu}\\
F_{\mu\nu}^{I-} &= -2\frac{\langle\bar{X}^I\rangle}{a} 
\sigma\star\tilde{F}_{\mu\nu}^- +l^I\bar{\sigma}\star\tilde{F}_{\mu\nu}^-
+r^I_{\ a}F_{\mu\nu}^-\star\tilde{\sigma}^a \\
\partial_\mu\Omega^I_i &=
\frac{l^I}{2}\varepsilon_{ik}\gamma^\rho\lambda^k\star\tilde{F}_{\rho\mu}^- 
+r^I_{\ a}\partial_\mu\lambda_i\star\tilde{\sigma}^a \\
\partial_\mu\bar{X}^I &= -\frac{\bar{l}^I}{2}F_{\mu\rho}^-\star\tilde{A}^\rho
+\bar{r}^I_{\ a}\partial_\mu\bar{\sigma}\star\tilde{\sigma}^a \;,
\end{aligned}
} 
\ee
with the composite fields \eqref{comp_fields_lin} expressed as
\be
\label{the_dictionary_composite}
\begin{aligned}
aT_{\mu\nu}^-&=-4\sigma\star\tilde{F}_{\mu\nu}^-  \;,\\ 
\mathcal{F}_{\mu\nu}^{I+}&=
{\bar l}^I {\sigma}\star\tilde{F}_{\mu\nu}^+ +
{\bar r}^I_{\ a}F_{\mu\nu}^+\star\tilde{\sigma}^a \;.
\end{aligned}
\ee
The dictionary parameters $l^I$ and  $r^I_{\ a}$ have to satisfy the relations
\be 
\label{constraints_lr}
\begin{aligned}
\langle N_{IJ}\bar{X}^I \rangle l^J&=0 \;, \\
\langle N_{IJ}\bar{X}^I \rangle r^J_{\ a}&=0 \;.
\end{aligned}
\ee

Thus, the double copy expressions \eqref{the_dictionary_unpack} provide a consistent on-shell double copy
dictionary for linearised ${\mathcal N}=2$ supergravity coupled to an arbitrary number of vector multiplets.

\subsection{Dictionary parameters}

Let us return to the double copy expression \eqref{dict_scalar_individual} for the scalar fields $X^I$.
At the linearised level, when expressing $X^I$ in terms of $z^A = X^A/ X^0$, we get, using \eqref{splitpXI},
\be
\partial_{\mu} X^I = \partial_{\mu} z^A \, \langle D_A X^I \rangle \;.
\label{XzA}
\ee
The double copy relation for the scalar fields $X^I$ becomes
\be
 \partial_{\mu} z^A \, \langle D_A X^I \rangle =  -\frac{l^I}{2}F_{\mu\rho}^+\star\tilde{A}^\rho
+ r^I_{\ a}\partial_\mu \left(\sigma \star {\tilde{\sigma}}^a \right) \;,
\ee
which suggests that $l^I$ and  $r^I_{\ a}$ are proportional to $ \langle D_A X^I \rangle $. This, in turn, is consistent 
with the constraint \eqref{constraints_lr} by virtue of \eqref{XNDX}.

\section{Double copy description of multi-centered BPS solutions}
Next, we apply the double copy dictionary \eqref{the_dictionary_unpack} to multi-centered
dyonic BPS black hole solutions in ${\mathcal N} =2$ supergravity coupled to an arbitrary number of vector multiplets.
We refer to \autoref{App:multiBPSbh} for a brief review of some of the features of these black hole solutions.

We work with Cartesian coordinates. The  linearised metric $g_{\mu \nu} = \eta_{\mu \nu} + h_{\mu \nu}$ is given in  \eqref{linelinmc}.
On the double copy side, we take the dictionary parameter $a$ to be real, and we express the 
metric fluctuation $h_{\mu \nu}$ as
\begin{equation}
a \, h_{\mu \nu} = A_{\mu} \star {\tilde A}_{\nu} + A_{\nu} \star {\tilde A}_{\mu} -  \left(A_{\alpha} \star {\tilde A}^{\alpha}
\right) \eta_{\mu \nu} \;.
\label{hdcprel}
\end{equation}
For the double copy ansatz we take, 
\begin{eqnarray}
\label{AtAdc}
A_{\mu}= \left(\frac{a \, Q}{r},- a \, \omega_1, - a \, \omega_2, -a \, \omega_3\right) \;\;\;,\;\;\; 
\tilde{A}_{\mu} = \left(\delta^{(4)}(x),0,0,0\right) \;,
\end{eqnarray}
which, when inserted into \eqref{hdcprel}, generates
the linearised metric fluctuation $h_{\mu \nu}$ given in \eqref{linelinmc}.\footnote{The Lorentz type condition $\partial_{\mu} \tilde{A}^{\mu}=0$  imposed in the derivation
of the double copy dictionary is, in fact, a special case of the more general constraint $\partial_{\mu} \left(
\varphi \star \tilde{A}^{\mu} \right) =0$, where $\varphi$ is an arbitrary on-shell field in the SYM sector, see
\cite{us}. We note that \eqref{AtAdc} satisfies this more general constraint.}

Given \eqref{AtAdc}, 
we compute
\be
F^-_{t j} = \frac12 \left(F_{tj} + \frac{i}{2} \varepsilon_{jkl} F^{kl} \right) = - \left( \frac{a \, Q}{2} 
\, \partial_j \frac{1}{r} + \frac{i}{2} a \, R_j(\omega) \right) \;\;\;,\;\;\; 
\tilde{F}_{tj}^- = - \frac12 \partial_j \delta^4(x) \;,
\label{FantisdRom}
\ee
with $R_j (\omega)$ defined in \eqref{Rom}. Next we recall that $F = d A$ satisfies the BPS relations
(see subsection 3.1 of \cite{us}) 
\be
F_{tj} = - 2 {\bar k}\, \partial_j {\rm Re} \; \sigma  \;\;\;,\;\;\; F_{jk} = -2 {\bar k} \, \varepsilon_{jkl} 
\partial^l {\rm Im} \, \sigma  \;\;\;,\;\;\; \partial_t \sigma  = 0 \;,
\ee
and hence $F^-_{t j} = -  {\bar k} \, \partial_j \sigma $. Then, comparing with \eqref{FantisdRom}, we infer 
\be
{\bar k} \, \partial_j  \sigma   = \frac{a \, Q}{2} 
\, \partial_j \frac{1}{r} + \frac{i}{2} a \, R_j(\omega) \;.
\label{sigQR}
\ee
This satisfies the equation of motion $\partial^j \partial_j   \sigma   = 0$ (in the absence of sources).

Using the above, we compute
\be
F^+_{t \nu} \star {\tilde A}^{\nu} = 0 \;\;\;,\;\;\; 
F^+_{j \nu} \star {\tilde A}^{\nu} = F_{tj}^+ = - k \, \partial_j \bar \sigma  \;.
\ee
The above equation yields a double copy expression for the scalar fields $X^I$ as 
\be
 \partial_t {X}^I = 0 \;\;\;,\;\;\; 
{\bar k} \, \partial_j {X}^I = \frac{{l}^I}{2} \partial_j \bar \sigma
+ {\bar k} \, {r}^I_{\ a} \,  \partial_j  \left(  {\sigma}\star\tilde{\sigma}^a \right) \;,
\label{dcXmulti}
\ee
while $F_{tj}^{I-}$ becomes
\be
\begin{aligned}
F_{tj}^{I-} &= -2\frac{\langle\bar{X}^I\rangle}{a} 
\sigma\star\tilde{F}_{tj}^- +l^I\bar{\sigma}\star\tilde{F}_{tj}^-
+r^I_{\ a}F_{tj}^-\star\tilde{\sigma}^a \\
&= \frac{\langle\bar{X}^I\rangle}{a} 
\partial_j \sigma - \frac{l^I}{2} \, \partial_j \bar{\sigma} -{\bar k} \, r^I_{\ a} \partial_j 
\left( \sigma \star\tilde{\sigma}^a \right) \;.
\end{aligned}
\ee
This we compare with the supergravity expression \eqref{sugraFtiasd2}. Making use of \eqref{dcXmulti}
and of \eqref{sigQR}, we find that both expressions match.

Using \eqref{Romc}, we infer from \eqref{sigQR},
\be
{\bar k} \, \partial_j  \sigma  = \frac{a \, (Q + i c) }{2} 
\, \partial_j \frac{1}{r} \;.
\label{sigQc}
\ee
For the scalar fields ${\tilde \sigma}^a$ we take
\be
 {\tilde \sigma}^a= \beta^a \,  \delta^{(4)} (x) \;,
 \label{valuesigtia}
\ee
with $\beta^a$ real constants.  This, together with \eqref{sigQc} and \eqref{AtAdc}, specifies
the double field configuration for multi-center BPS black holes. Inserting \eqref{valuesigtia}
in \eqref{dcXmulti} gives
\be
 \partial_j {X}^I = \frac{{l}^I}{2} \, k \, \partial_j \bar \sigma
+{r}^I_{\ a} \, \beta^a \, \partial_j   {\sigma} \;.
\label{scalarsig1}
\ee
The scalar field fluctuations $X^I$ take the form $X^I = \Sigma^I/r$, see \autoref{App:multiBPSbh}.
Thus, both $\partial_j {X}^I$ and $ {\bar k}  \, \partial_j  \sigma  $ behave as $\partial_j (1/r)$.
Then, from \eqref{scalarsig1}, we obtain
\be
\Sigma^I = \frac{l^I}{4} \, a (Q - ic) + \frac{k \, {r}^I_{\ a} \, \beta^a}{2} 
\, a (Q + ic) \;.
\label{sigQc2}
\ee
We note that we may simplify \eqref{sigQc2} by setting $\beta^a =0$, in which case
$\tilde{\sigma}^a =0$. This results in the determination of the $l^I$ as
\be
l^I = \frac{4 \Sigma^I}{ a (Q - ic)} = \frac{4  \mu^A \, \langle D_A X^I \rangle }{ a (Q - ic)} \;,
\ee
where we used \eqref{splitpXI}, and
with $\Sigma^I$ and $\mu^A$ determined as in \autoref{App:multiBPSbh}.

Thus, we conclude that in the weak field approximation, dyonic multi-centered BPS black hole solutions have a double copy description, based on \eqref{the_dictionary_unpack},
in terms of a charged BPS field theory configuration specified by \eqref{AtAdc}.

At this point, we recall (see the discussion in \autoref{sect-introduction}) that we have omitted the dependence
on the spectator field for notational simplicity,
\be
\varphi \star  {\tilde \varphi} \equiv \varphi^{\alpha} \star \phi_{{\alpha} \tilde \alpha} \star {\tilde \varphi}^{\tilde \alpha} \;.
\ee
This dependence can be reinstated by taking, as we did in \cite{us}, $A_\mu^\alpha=A_\mu c^\alpha$, $\tilde{A}_\mu^{\tilde{\alpha}}=\tilde{A}_\mu \tilde{c}^{\tilde{\alpha}}$ and $\phi_{\alpha\tilde{\alpha}}=V_{\alpha\tilde{\alpha}}\delta^4(x)$, with constants $c^\alpha,\tilde{c}^{\tilde{\alpha}},V_{\alpha\tilde{\alpha}}$ normalised to $c^\alpha V_{\alpha\tilde{\alpha}}\tilde{c}^{\tilde{\alpha}}=1$.

Additionally, in \autoref{sect-introduction} we stated that the spectator field is to be thought of as the convolution inverse of a bi-adjoint scalar field. The double copy prescription then requires us to consider solutions for the bi-adjoint scalar theory (which we denoted by $\Sigma_{\alpha\tilde{\alpha}}$) and plug their convolution inverse into the double copy (via $\phi_{\alpha\tilde{\alpha}}=[\Sigma^{-1}]_{\alpha\tilde{\alpha}}$). The spectator we have chosen in this case ($\phi_{\alpha\tilde{\alpha}}=V_{\alpha\tilde{\alpha}}\delta^4(x)$) is to be interpreted as the inverse of a point-source\footnote{We recall that the delta function is its own convolution inverse $\delta\star\delta=\delta$.}  $\Sigma^{\alpha\tilde{\alpha}}=W^{\alpha\tilde{\alpha}}\delta^4(x)$, such that $W^{\alpha\tilde{\alpha}}V_{\alpha\tilde{\alpha}}=1$.

Incidentally, the above discussion immediately points to the non-uniqueness of the double-copy. 
Instead
of choosing a point-source for $\tilde{A}_\mu^{\tilde{\alpha}}$ (as given in \eqref{AtAdc}), 
one may choose $\tilde{A}_\mu^{\tilde{\alpha}}$ differently, for instance 
$\tilde{A}_\mu^{\tilde{\alpha}}= (\tilde{c}^{\tilde{\alpha}} \, B(r),0,0,0)$, together with a spectator field 
$\phi_{\alpha\tilde{\alpha}}=V_{\alpha\tilde{\alpha}} \left[B(r) \right]^{-1}$, to recover the same 
supergravity solution. In principle this allows for a more symmetric treatment of the two YM gauge fields
in \eqref{AtAdc}.

\section{Conclusions}

In this note we have constructed a weak-field double copy dictionary for on-shell configurations in four-dimensional ${\cal N}=2$ supergravity theories consisting of gravity coupled to an arbitrary number of vector multiplets and described multi-centered BPS black holes using this double copy framework. The basic double copy dictionary relates a gravitational on-shell field $\varphi_G$ to a convolution of on-shell field theory fields $\varphi$ and 
$\tilde{\varphi}$ as 
\begin{equation}\label{prop}
\varphi_G = \int d^4 y\, \varphi(y) \tilde{\varphi}(x-y)\,,
\end{equation}
where the spectator field is suppressed for notational simplicity .
In  holographic gauge-gravity duality, a bulk gravitational operator is constructed from 
the corresponding holographic boundary operator by convolving it with a so-called non-local smearing function, which acts as a bulk-boundary propagator. Hence, the holographic  relation obtained  between the bulk and boundary quantities is similar to \eqref{prop} with the second factor in the integrand, $\tilde{\varphi}(x-y)$,  playing the role of the bulk-boundary propagator. This might indicate a potential handle on the long-standing problem of developing a holographic duality for asymptotically flat backgrounds.
In parallel to the usual holographic dictionary, the local field in the convolution integral is a normalizable fluctuation of the 'bulk' field. One essential check on this potential holographic perspective  is to investigate 
whether the BMS asymptotic symmetry group  at null infinity \cite{Bondi:1962px}\footnote{These correspond to diffeomorphisms that preserve the asymptotic metric but are not isometries of the full spacetime, and hence  only encode information about the spacetime in the weak-field approximation.} corresponds, via the double copy dictionary, to an asymptotic symmetry group of the SYM field theory sector that appears in the double copy dictionary.
A positive answer to this question will set the ground for exploring whether the states in the gravitational theory are organized in terms of symmetries of the  gauge theory with appropriate gauge-gravity propagators. Strictly speaking, this will  not be a holographic duality as the gauge field lives in the same dimension of spacetime as the gravitational configurations in question. But it should be viewed as a putative gauge-gravity correspondence, which, if established consistently, will encode gravitational configurations in terms of gauge degrees of freedom in the same spirit as the holographic AdS/CFT correspondence. 
These open questions present promising avenues of research which will be pursued in the near future.


\subsection*{Acknowledgements}

\noindent
We would like to thank Leron Borsten  for fruitful discussions.  
This work was supported by FCT/Portugal through a CAMGSD post-doctoral fellowship (S. Nagy)  and through FCT fellowship SFRH/BPD/101955/2014 (S. Nampuri).
This work was also supported by the COST action MP1210
"The String Theory Universe". 
\appendix

\section{Special geometry} \label{App:Kahlerconnec}

We review a few elements of special geometry \cite{deWit:1984rvr,Strominger:1990pd}.

The Einstein frame constraint
\be 
i \left[ {\bar X}^I F_I - X^I \, {\bar F}_I \right] = 1 \;\;\;,\;\;\; I=0, \dots , n_V \;,
\ee
can be solved by setting 
\be
X^0 = e^{K/2} \, X^0 (z)  \;\;\;,\;\;\; X^A = e^{K/2} \, X^0 (z) \, z^A  \;\;\;,\;\;\; A=1, \dots , n_V \;,
\label{X0Az}
\ee
where
\be
K = {\hat K}  - \ln X^0(z) - \ln {\bar X}^0 (\bar z) \;,
\ee
with
\be
{\hat K} = - \ln \left[ - {\bar z}^I \, z^J \, N_{IJ} \right] 
\label{hatKz}
\ee
and
\be
N_{IJ} = -i \left(F_{IJ} - {\bar F}_{\bar I \bar J} \right) \;,
\ee
and we defined
\be
z^I = \frac{X^I}{X^0} \;.
\label{zIdef}
\ee
Observe that
\be
|X^0|^2 = e^{\hat K} \;.
\label{X0hatK}
\ee
The holomorphic transformation
\be
X^0(z) \rightarrow e^{- f(z)} \, X^0(z) 
\label{kaehlX0}
\ee
induces a K\"ahler transformation,
\be
K \rightarrow K + f + \bar f \;.
\ee
Note that $z^I$ and ${\hat K}$ are inert under this transformation, while the $X^I$ transform 
with a $U(1)$ phase,
\be
X^I \rightarrow e^{-(f - \bar f)/2} \, X^I \;.
\ee
The
$U(1)$ connection \eqref{u1a}, when expressed in terms of $z^A$ and ${\bar z}^A$, equals
\be
a_{\mu} = \frac{i}{2} \left( \partial_{\mu} z^A \, \partial_A K - \partial_{\mu} {\bar z}^A \, \partial_{\bar A} K \right) \;\;\;,\;\;\;
 \partial_A = \frac{\partial}{\partial z^A} \;\;\;,\;\;\; \partial_{\bar A} = \frac{\partial}{\partial {\bar z}^A} \;.
\ee
K\"ahler covariant derivatives of $X^I$ are defined by 
\begin{eqnarray}
D_A X^I &=& \partial_A X^I + \frac12 \, \left( \partial_A K \right)
\, X^I \;,
\nonumber\\
{D}_{\bar A} X^I &=& \partial_{\bar A} X^I - \frac12 \, \left( \partial_{\bar A} K \right)
\, X^I \;.
\end{eqnarray}
Using \eqref{X0Az}, it follows that
\be
{D}_{\bar A} X^I = 0 \;.
\ee
Then we obtain
\be
\partial_{\mu} X^I 
= \partial_{\mu} z^A \, D_A X^I + i a_{\mu} \, X^I\;.
\label{derXkahla}
\ee

At the linearised level, the $U(1)$ connection vanishes, as follows.
Consider the linearised S-supersymmetry gauge fixing condition \eqref{s_super_lin}.
Taking a Q-supersymmetry of \eqref{s_super_lin} results in
\be 
\gamma^\mu\langle X^I \, N_{IJ} \rangle\partial_\mu\bar{X}^J\epsilon^i +\frac{1}{4}\gamma^{\mu\nu}
\langle X^I \, N_{IJ} \rangle\mathcal{F}_{\mu\nu}^{J+}\varepsilon^{ij}\epsilon_j=0 \;.
\label{susySom}
\ee
Using \eqref{comp_fields_lin}, one verifies that the second term vanishes,
and hence from \eqref{susySom} one infers
\be 
\langle X^I \, N_{IJ} \rangle\partial_\mu\bar{X}^J=0 \;.
\label{XNderX}
\ee
Since, at the linearised level, the
$U(1)$ connection \eqref{u1a} reads
\be
a_{\mu} =-\frac{i}{2}\langle X^I N_{IJ}\rangle\partial_\mu\bar{X}^J+c.c. \;,
\ee
it vanishes by virtue of \eqref{XNderX}. Hence, at the linearised level, \eqref{derXkahla} becomes
\be
\partial_{\mu} X^I = \partial_{\mu} z^A \, \langle D_A X^I \rangle \;.
\label{splitpXI}
\ee

Next, let us compute $D_A X^I$,
\begin{eqnarray}
D_A X^0 &=& \partial_A {\hat K} \, X^0 \:, \nonumber\\
D_A X^B &=& \delta_A^B \, X^0 +  \partial_A {\hat K} \, X^B \;,
\end{eqnarray}
where we used \eqref{X0Az}.  
Using these expressions, we compute
\be
{\bar X}^I N_{IJ} D_A X^J = {\bar X}^I N_{IJ} X^J \, \partial_A {\hat K} + {\bar X}^I N_{IA} X^0 = 
- \partial_A {\hat K} + {\bar X}^I N_{IA} X^0 \;.
\label{bXNDX}
\ee
Using \eqref{hatKz}, we infer
\begin{equation}
\frac{\partial}{\partial z^A} e^{- \hat K} = - \partial_A {\hat K} \,  e^{- \hat K} = - {\bar z}^I N_{IA} \;,
\ee
where we used the special geometry relation $F_{IJK} X^K =0$. Hence
\be
\partial_A {\hat K} = {\bar z}^I N_{IA} \,  e^{\hat K} = {\bar X}^I N_{IA} X^0 \;,
\ee
where we used \eqref{X0hatK}. It follows that \eqref{bXNDX} vanishes, 
\be
{\bar X}^I N_{IJ} D_A X^J = 0 \;.
\label{XNDX}
\ee

\section{Double copy expressions for the gravitini}  \label{App:alterndcferm}

Here we prove that the double copy expression \eqref{expressiongrav} for the gravitini 
\be 
a\psi_{\mu\nu}^i =
\varepsilon^{ij}  \lambda_j \star  \tilde{F}^-_{\mu \nu}
\ee
can be brought into the form
\eqref{gravidc}
\be 
a\psi_{\mu\nu}^i=
\varepsilon^{ij}\gamma^\rho\gamma_{[\nu}\lambda_j \star \partial_{\mu]}\tilde{A}_\rho,
\ee
by making use of the equation of motion for $\lambda_j$ and the Lorentz gauge condition $\partial_{\mu} {\tilde A}^{\mu} =0$. We start with 
\be
\begin{aligned}
a\psi_{\mu\nu}^i&=\varepsilon^{ij}\lambda_j \star\tilde{F}_{\mu\nu}^-\\
&=\varepsilon^{ij}\lambda_j \star\frac{1}{2}\left[\tilde{F}_{\mu\nu}
+\frac{i}{2}\varepsilon_{\mu\nu\alpha\beta}\tilde{F}^{\alpha\beta}
\right]\\
&=\varepsilon^{ij}\left[\encircle{A}+\encircle{B}\right] \;.
\end{aligned} 
\ee
Now we use $-i\varepsilon_{\mu\nu\alpha\beta}=\gamma_{\mu\nu\alpha\beta}\gamma_5$ and $\gamma_5\lambda_j=\lambda_j$ to write
\be
\encircle{B}=-\frac{1}{2}\gamma_{\mu\nu\alpha\beta}\lambda_j\star
\partial^\alpha\tilde{A}^\beta \;.
\ee
Then we use $\gamma_{\mu\nu\alpha\beta}=\gamma_{\alpha\beta\mu\nu}=\frac{1}{2}[\gamma_\alpha,\gamma_{\beta\mu\nu}]$, together with the equation of motion for $\lambda_j$ to get
\be
\begin{aligned}
\encircle{B}&=
-\frac{1}{4}\gamma_\alpha\gamma_{\beta\mu\nu}\partial^\alpha\lambda_j\star
\tilde{A}^\beta\\
&=-\frac{1}{8}\gamma_\alpha\gamma_\beta\gamma_{\mu\nu}\partial^\alpha\lambda_j\star
\tilde{A}^\beta
-\frac{1}{8}\gamma_\alpha\gamma_{\mu\nu}\gamma_\beta\partial^\alpha\lambda_j\star
\tilde{A}^\beta\\
&=\encircle{$B_1$}+\encircle{$B_2$} \;.
\end{aligned}
\ee
Then
\be
\begin{aligned}
\encircle{$B_1$}&=-\frac{1}{16}\gamma_\alpha\gamma_\beta\gamma_\mu\gamma_\nu\partial^\alpha\lambda_j
\star\tilde{A}^\beta-(\mu\leftrightarrow\nu)\\
&=0+\frac{1}{16}\gamma_\beta\gamma_\alpha\gamma_\mu\gamma_\nu\partial^\alpha\lambda_j
\star\tilde{A}^\beta-(\mu\leftrightarrow\nu)\\
&=\frac{1}{8}\gamma_\beta\gamma_\nu\partial_\mu\lambda_j\star\tilde{A}^\beta
-\frac{1}{16}\gamma_\beta\gamma_\mu\gamma_\alpha\gamma_\nu\partial^\alpha\lambda_j\star\tilde{A}^\beta
-(\mu\leftrightarrow\nu)\\
&=\frac{1}{8}\gamma_\beta\gamma_\nu\partial_\mu\lambda_j\star\tilde{A}^\beta
-\frac{1}{8}\gamma_\beta\gamma_\mu\partial_\nu\lambda_j\star\tilde{A}^\beta
-(\mu\leftrightarrow\nu)\\
&=\frac{1}{2}\gamma^\rho\gamma_{[\nu}\partial_{\mu]}\lambda_j\star\tilde{A}_\rho \;,
\end{aligned} 
\ee
and the second term is
\be
\begin{aligned}
\encircle{$B_2$}
&=-\frac{1}{16}\gamma_\alpha\gamma_\mu\gamma_\nu\gamma_\beta\partial^\alpha\lambda_j\star\tilde{A}^\beta
-(\mu\leftrightarrow\nu)\\
&=-\frac{1}{8}\gamma_\alpha\gamma_\mu\partial^\alpha\lambda_j\star\tilde{A}_\nu
+\frac{1}{16}\gamma_\alpha\gamma_\mu\gamma_\beta\gamma_\nu\partial^\alpha\lambda_j\star\tilde{A}^\beta
-(\mu\leftrightarrow\nu)\\
&=-\frac{1}{4}\lambda_j\star\partial_\mu\tilde{A}_\nu+\frac{1}{8}\gamma_\alpha\gamma_\nu
\partial^\alpha\lambda_j\star\tilde{A}_\mu-(\mu\leftrightarrow\nu)+\encircle{$B_1$}\\
&=-\frac{1}{2}\lambda_j\star\tilde{F}_{\mu\nu}+\encircle{$B_1$}\\
&=-\encircle{A}+\encircle{$B_1$} \;.
\end{aligned} 
\ee
Thus we obtain
\be
\begin{aligned}
a\psi_{\mu\nu}^i&=\varepsilon^{ij}\left[\encircle{A}+\encircle{B}\right]\\
&=2\varepsilon^{ij}\encircle{$B_1$}\\
&=\varepsilon^{ij}\gamma^\rho\gamma_{[\nu}\partial_{\mu]}\lambda_j\star\tilde{A}_\rho \;,
\end{aligned} 
\ee
which matches \eqref{gravidc}.

\section{Dictionary derivation} \label{App:dderiva}

In subsection \ref{onshelldcdict} we determined the double copy expression for $\psi_{\mu \nu}^i$ and $\partial_{\mu} 
\Omega^{Ii}$. Here, we first derive the dictionary for the other field strengths appearing in the dictionary
\eqref{the_big_dictionary}. 
We will use the notation $\delta_Q \varphi (\Sigma)$ to indicate that we are only considering terms in the
variation of a field $\varphi$ that are proportional to $\Sigma$. We refer to appendix A of \cite{us}
for a summary of the conventions we use.

We begin by working out 
the
double copy expressions for
$T_{\mu\nu}^-$ and $\mathcal{F}_{\mu\nu}^{I+}$, defined in \eqref{comp_fields_lin}. 
The derivation of the double copy expression for $T_{\mu\nu}^-$ is analogous to the one given in 
appendix D of \cite{us}, so we refrain from repeating it here. The double copy expression for 
$\mathcal{F}_{\mu\nu}^{I+}$ is obtained from the supersymmetry variations of the gaugini $\Omega^{Ii}$.
Using \eqref{all_sugra_SUSY}, we have
\be
\label{calf_from_sugra}
2b_I\delta_Q\partial_\mu\Omega^{Ii}(\mathcal{F})
=\frac{1}{2}\gamma^{\alpha\beta}\partial_\mu(b_I\mathcal{F}_{\alpha\beta}^{I+}) \varepsilon^{ij}\epsilon_j \;.
\ee
We match this with the supersymmetry variation on the double copy side.  Using 
the convolution property \eqref{shift-der}, we obtain
\be
\begin{aligned}
2b_I\delta_Q\partial_\mu\Omega^{Ii}(\mathcal{F})&=
\varepsilon^{ij}\gamma^\rho\delta_Q\lambda_j(\sigma)\star\partial_\mu\tilde{A}_\rho
+2c_a\partial_\mu\delta_Q\lambda^i(F)\star\tilde{\sigma}^a\\
&=\gamma^\rho\gamma^\alpha\partial_\alpha\sigma\star\partial_\mu\tilde{A}_\rho
\varepsilon^{ij}\epsilon_j
+\frac{c_a}{2}\partial_\mu(\gamma^{\alpha\beta}F_{\alpha\beta}^+)\star\tilde{\sigma}^a
\varepsilon^{ij}\epsilon_j\\
&=\gamma^\rho\gamma^\alpha\partial_\mu(\sigma\star\partial_\alpha\tilde{A}_\rho)
\varepsilon^{ij}\epsilon_j+\frac{c_a}{2}\partial_\mu(\gamma^{\alpha\beta}F_{\alpha\beta}^+)\star\tilde{\sigma}^a
\varepsilon^{ij}\epsilon_j \;,
\end{aligned} 
\ee
Making use of $\partial^\rho\tilde{A}_\rho=0$, we obtain
\be
\label{calf_from_SYM}
\begin{aligned}
2b_I\delta_Q\partial_\mu\Omega^{Ii}(\mathcal{F})&=-\frac{1}{2}\gamma^{\rho\alpha}\partial_\mu(\sigma\star\tilde{F}_{\rho\alpha})
\varepsilon^{ij}\epsilon_j
+\frac{c_a}{2}\partial_\mu(\gamma^{\alpha\beta}F_{\alpha\beta}^+)\star\tilde{\sigma}^a
\varepsilon^{ij}\epsilon_j\\ 
&=-\frac{1}{2}\gamma^{\alpha\beta}\partial_\mu(\sigma\star\tilde{F}_{\alpha\beta}^+
-c_a F_{\alpha\beta}^+\star\tilde{\sigma}^a)
\varepsilon^{ij}\epsilon_j \;.
\end{aligned}
\ee
Then, comparing \eqref{calf_from_sugra} and \eqref{calf_from_SYM}, we find
\be 
b_I\mathcal{F}_{\mu\nu}^{I+}=-\sigma\star\tilde{F}_{\mu\nu}^+
+c_a F_{\alpha\beta}^+\star\tilde{\sigma}^a \;.
\ee

Next, we derive the double copy dictionary for the scalars $X^I$.
We make use of the supersymmetry transformation of the gaugino \eqref{all_sugra_SUSY} to write
\be
\label{app_scalar_from_sugra}
2b_I\delta_Q\partial_\mu\Omega^{Ii}(X)=2b_I\gamma^\rho\partial_\mu(\partial_\rho\bar{X}^I)\epsilon^i \;.
\ee   
We will compare this with the supersymmetry variation on the double copy side,
\be
\begin{aligned}
2b_I\delta_Q\partial_\mu\Omega^{Ii}(X)&=
\varepsilon^{ij}\gamma^\rho\delta_Q\lambda_j(F)\star\partial_\mu\tilde{A}_\rho
+2c_a\partial_\mu\delta_Q\lambda^i(\sigma)\star\tilde{\sigma}^a\\
&=-\frac{1}{4}\gamma_\rho\gamma^{\alpha\beta}F_{\alpha\beta}^- \star\partial_\mu\tilde{A}^\rho\epsilon^i
+2c_a\gamma^\rho\partial_\mu\partial_\rho\bar{\sigma}\star\tilde{\sigma}^a\epsilon^i \;.
\end{aligned} 
\ee
Using the relation \eqref{app_anti_self_contract},
we obtain
\be
\label{app_scalar_from_SYM}
\begin{aligned}
2b_I\delta_Q\partial_\mu\Omega^{Ii}(X)&=\gamma^\alpha F_{\alpha\rho}^-\star\partial_\mu\tilde{A}^\rho\epsilon^i
+2c_a\gamma^\rho\partial_\mu\partial_\rho\bar{\sigma}\star\tilde{\sigma}^a\epsilon^i\\
&=\gamma^\rho\partial_\mu(F_{\rho\nu}^-\star\tilde{A}^\nu+2c_a\partial_\rho\bar{\sigma}\star\tilde{\sigma}^a)\epsilon^i
\;.
\end{aligned} 
\ee
Then, comparing \eqref{app_scalar_from_sugra} and \eqref{app_scalar_from_SYM}, we read off the dictionary for the scalar
$X^I$, 
\be 
b_I\partial_\mu\bar{X}^I=\frac{1}{2}F_{\mu\rho}^-\star\tilde{A}^\rho+c_a\partial_\mu\bar{\sigma}\star\tilde{\sigma}^a\;.
\ee
Finally, we note that the derivation of the double copy expression for the Riemann tensor is analogous to the one given in 
appendix D of \cite{us}, so we refrain from repeating it here. We have thus derived the double copy
expressions in the dictionary
\eqref{the_big_dictionary}. 

In deriving the dictionaries \eqref{the_big_dictionary} and \eqref{the_dictionary_unpack}, we used the supersymmetry
variation of the double copy expressions of some of the fields involved. We now check 
that both sides of the expressions in the dictionary 
\eqref{the_dictionary_unpack}
transform consistently under supersymmetry.

To this end, we will 
use the following supergravity relations that hold at the linearized level,
\be
\label{sugrarelgg}
\begin{aligned}
\delta_Q F_{\mu\nu}^{I-} &=\varepsilon^{ij}\bar{\epsilon}_i\gamma_{[\nu}\partial_{\mu]}\Omega^I_j
+\varepsilon_{ij}\bar{\epsilon}^i\psi_{\mu\nu}^j\langle \bar{X}^I\rangle \;, \\
\delta_Q T_{\mu\nu}^{-} &= 2 \varepsilon_{ij} {\bar \epsilon}^i \psi^j_{\mu \nu} \;, \\
\delta_Q\mathcal{F}_{\mu\nu}^{I+}
&= \varepsilon_{ij}\bar{\epsilon}^i\gamma_{[\nu}\partial_{\mu]}\Omega^{Ij} \;.
\end{aligned}
\ee   
Let us first consider $\delta_Q T_{\mu\nu}^{-}$,
\be
\label{T_min_from_squaring}
\begin{aligned}
a\delta_Q T_{\mu\nu}^{-}&=-4\delta_Q\sigma\star\tilde{F}_{\mu\nu}^- \\
&=-2\bar{\epsilon}^i\lambda_i\star\tilde{F}_{\mu\nu}^- \\
&=-\bar{\epsilon}^i\lambda_i\star\tilde{F}_{\mu\nu}
+\ ^*(\bar{\epsilon}^i\lambda_i\star\tilde{F}_{\mu\nu}) \;.
\end{aligned} 
\ee
Now we make use of \eqref{dict_ansatz} to write
\be
\label{lambda_F_munu}
\bar{\epsilon}^i\lambda_i\star\tilde{F}_{\mu\nu}
=-a\varepsilon_{ij}\bar{\epsilon}^i\psi_{\mu\nu}^j
-2b_I\varepsilon_{ij}\bar{\epsilon}^i\gamma_{[\nu}\partial_{\mu]}\Omega^{Ij} 
+2c_a\varepsilon_{ij}\bar{\epsilon}^i\gamma_{[\nu}\partial_{\mu]}\lambda^j
\star\tilde{\sigma}^a \;.
\ee 
One can prove that the relations
\be
\label{useful_spinors}
\begin{aligned}
^*(\varepsilon^{ij}\bar{\epsilon}_i\gamma_{[\nu}\partial_{\mu]}\chi_j)&=
-\varepsilon^{ij}\bar{\epsilon}_i\gamma_{[\nu}\partial_{\mu]}\chi_j \;, \\
^*(\varepsilon_{ij}\bar{\epsilon}^i\gamma_{[\nu}\partial_{\mu]}\chi^j)&=
\varepsilon_{ij}\bar{\epsilon}^i\gamma_{[\nu}\partial_{\mu]}\chi^j \;,
\end{aligned}
\ee
and
\be
\label{useful_gravitini}
\begin{aligned}
^*(\varepsilon^{ij}\bar{\epsilon}_i\psi_{\mu\nu j})
&=\varepsilon^{ij}\bar{\epsilon}_i\psi_{\mu\nu j} \;, \\
^*(\varepsilon_{ij}\bar{\epsilon}^i\psi_{\mu\nu}^j)
&=-\varepsilon_{ij}\bar{\epsilon}^i\psi_{\mu\nu}^j \;,
\end{aligned} 
\ee
hold on-shell at the linearised level. We use them to obtain
\be
\label{dual_lambda_F_munu}
^*(\bar{\epsilon}^i\lambda_i\star\tilde{F}_{\mu\nu})=
a\varepsilon_{ij}\bar{\epsilon}^i\psi_{\mu\nu}^j 
-2b_I\varepsilon_{ij}\bar{\epsilon}^i\gamma_{[\nu}\partial_{\mu]}\Omega^{Ij} 
+2c_a\varepsilon_{ij}\bar{\epsilon}^i\gamma_{[\nu}\partial_{\mu]}\lambda^j
\star\tilde{\sigma}^a \;,
\ee
and thus
\be
\label{final_T_min}
a\delta_Q T_{\mu\nu}^{-}=2a\varepsilon_{ij}\bar{\epsilon}^i\psi_{\mu\nu}^j  \;,
\ee
which matches the supergravity relation \eqref{sugrarelgg}.

Next, we check the supersymmetry variation of 
$b_I \mathcal{F}^{I+}_{\mu \nu}$. Using \eqref{sugrarelgg}, we infer
\be
\label{cFsug}
b_I \delta_Q\mathcal{F}_{\mu\nu}^{I+}
=b_I \varepsilon_{ij}\bar{\epsilon}^i\gamma_{[\nu}\partial_{\mu]}\Omega^{Ij} \;,
\ee
while from the double copy side we have
\be
\label{dccF}
b_I\delta_Q\mathcal{F}_{\mu\nu}^{I+}
= -\delta_Q\sigma\star\tilde{F}_{\mu\nu}^+
+c_a \delta_Q F_{\mu\nu}^+\star\tilde{\sigma}^a \;.
\ee
We obtain
\be
\begin{aligned}
-\delta_Q\sigma\star\tilde{F}_{\mu\nu}^+
&=-\frac{1}{2}\bar{\epsilon}^i\lambda_i\star\tilde{F}_{\mu\nu}^+\\
&=-\frac{1}{4}\bar{\epsilon}^i\lambda_i\star\tilde{F}_{\mu\nu}
-\frac{1}{4}\ ^*(\bar{\epsilon}^i\lambda_i\star\tilde{F}_{\mu\nu})\\
&=b_I\varepsilon_{ij}\bar{\epsilon}^i\gamma_{[\nu}\partial_{\mu]}\Omega^{Ij}
-c_a\varepsilon_{ij}\bar{\epsilon}^i\gamma_{[\nu}\partial_{\mu]}\lambda^i
\star\tilde{\sigma}^a
\end{aligned} 
\ee
where we used \eqref{lambda_F_munu} and \eqref{dual_lambda_F_munu}. Using
\be 
\label{F_plus_yangmills}
\delta_Q F_{\mu\nu}^+\star\tilde{\sigma}^a
=\varepsilon_{ij}\bar{\epsilon}^i\gamma_{[\nu}\partial_{\mu]}\lambda^j
\star\tilde{\sigma}^a \;,
\ee
we obtain for \eqref{dccF},
\be 
b_I\delta_Q\mathcal{F}_{\mu\nu}^{I+}
=b_I\varepsilon_{ij}\bar{\epsilon}^i\gamma_{[\nu}\partial_{\mu]}\Omega^{Ij} \;,
\ee
which matches \eqref{cFsug}. 

Next, we consider the supersymmetry variation of $X^I$. On the supergravity side we have
\be 
\label{scsug}
\delta_Q \partial_\mu X^I=\frac{1}{2}\bar{\epsilon}^i\partial_\mu\Omega_i^I \;,
\ee
whereas on the double copy side, using \eqref{dict_scalar_individual}, we get
\be  
\begin{aligned}
\delta_Q\partial_\mu X^I
&= -\frac{l^I}{2}\delta_QF_{\mu\rho}^+\star\tilde{A}^\rho
+r^I_{\ a}\delta_Q\partial_\mu\sigma\star\tilde{\sigma}^a\\
&= -\frac{l^I}{2}\varepsilon_{ij}\bar{\epsilon}^i\gamma_{[\rho}\partial_{\mu]}\lambda^j
\star\tilde{A}^\rho
+\frac{1}{2}r^I_{\ a}\partial_\mu\bar{\epsilon}^i\lambda_i\star\tilde{\sigma}^a\\
&= -\frac{l^I}{4}\varepsilon_{ij}\bar{\epsilon}^i\gamma_{\rho}\partial_{\mu}\lambda^j
\star\tilde{A}^\rho
+\frac{1}{2}r^I_{\ a}\partial_\mu\bar{\epsilon}^i\lambda_i\star\tilde{\sigma}^a\\
&=\frac{1}{2}\bar{\epsilon}^i\partial_\mu\Omega_i^I \;,
\end{aligned}
\ee
where we used \eqref{F_plus_yangmills} and $\partial_{\rho} \tilde{A}^\rho =0$. We thus have agreement with 
\eqref{scsug}.

Next, we turn to the supersymmetry variation of the gaugini. On the supergravity side we have
\be 
\label{omsug}
\delta_Q\partial_\mu\Omega^I_i({\cal F})=\frac{1}{4}\gamma^{\rho\sigma}\partial_\mu
\mathcal{F}_{\rho\sigma}^{I-}\varepsilon_{ij}\epsilon^j \;,	
\ee
while from the double copy we get,
\be
\begin{aligned}
\delta_Q\partial_\mu\Omega^I_i({\cal F})&=-\frac{l^I}{2}\varepsilon_{ij}\gamma^\rho\delta_Q\lambda^j(\sigma)
\star\partial_\mu\tilde{A}_\rho
+r^I_{\ a}\delta_Q\lambda_i(F)\star\partial_\mu\tilde{\sigma}^a\\
&=-\frac{l^I}{2}\varepsilon_{ij}\gamma^\rho\gamma^\sigma\partial_\sigma
\bar{\sigma}\star\partial_\mu\tilde{A}_\rho\epsilon^j
+\frac{1}{4}r^I_{\ a}\gamma^{\rho\sigma}F_{\rho\sigma}^-
\star\partial_\mu\tilde{\sigma}^a\varepsilon_{ij}\epsilon^j\\
&=\frac{l^I}{4}\gamma^{\rho\sigma}\partial_\mu\bar{\sigma}
\star\tilde{F}_{\rho\sigma}^-\varepsilon_{ij}\epsilon^j
+\frac{1}{4}r^I_{\ a}\gamma^{\rho\sigma}F_{\rho\sigma}^-
\star\partial_\mu\tilde{\sigma}^a\varepsilon_{ij}\epsilon^j\\
&=\frac{1}{4}\gamma^{\rho\sigma}\partial_\mu\left[
l^I\bar{\sigma}\star\tilde{F}_{\rho\sigma}^-
+r^I_{\ a}F_{\rho\sigma}^-\star\tilde{\sigma}^a
\right]\varepsilon_{ij}\epsilon^j\\
&=\frac{1}{4}\gamma^{\rho\sigma}\partial_\mu\left[ 
F_{\rho\sigma}^{I-} -k^I\sigma\star\tilde{F}_{\rho\sigma}^-
\right]\varepsilon_{ij}\epsilon^j \;,
\end{aligned}
\ee 
where to get to the last line we used \eqref{ansatz_ind_F}. Next we recall that $p^I=-2\frac{\langle\bar{X}^I\rangle}{a}$ to obtain
\be
\begin{aligned}
\delta_Q\partial_\mu\Omega^I_i({\cal F})&=\frac{1}{4}\gamma^{\rho\sigma}\partial_\mu\left[ 
F_{\rho\sigma}^- -\frac{1}{2}\langle\bar{X}^I\rangle T_{\rho\sigma}^-
\right]\varepsilon_{ij}\epsilon^j\\
&=\frac{1}{4}\gamma^{\rho\sigma}\partial_\mu
\mathcal{F}_{\rho\sigma}^{I-}\varepsilon_{ij}\epsilon^j \;,
\end{aligned} 
\ee
which matches \eqref{omsug}.

Finally, we remark that both sides of the double copy expression for the Riemann tensor transform consistently under
supersymmetry.

\section{Multi-centered dyonic BPS black hole solutions} \label{App:multiBPSbh}
In Cartesian coordinates, multi-centered BPS black hole solutions are described by the line element,
\begin{equation}
ds^2= -e^{2g} (dt+ \omega_i \, dx^i)^2 + e^{-2g} dx^i dx^i \;,
\end{equation}
with $i=1,2,3$. Here, both $g$ and $\vec{\omega}$ are independent of $t$.
We define
\be
R(\omega)_i = \varepsilon_{i}{}^{jk} \,  \partial_j \omega_k \;.
\label{Rom}
\ee
The warp factor $g(x^i)$ is 
determined by
\be 
e^{-2g}=i(\bar{Y}^IF_I (Y) -Y^I \bar{F}_I (\bar Y)) \;.
\label{gY}
\ee
Here $F_I(Y)= \partial F(Y)/\partial Y^I$, with
the $Y^I$ defined by \cite{LopesCardoso:2000qm}
\be
Y^I=e^{-g} X^I\bar{k} \;,
\label{YX}
\ee
where $k$ denotes a $U(1)$  compensating phase.
The $Y^I$ 
are determined through the attractor mechanism \cite{Ferrara:1996dd} by the attractor equations
\be
\label{attractor_equations}
\begin{aligned}
Y^I-\bar{Y}^I&=i H^I \;, \\
F_I(Y)-\bar{F}_{\bar I} (\bar{Y})&=iH_I \;,
\end{aligned} 
\ee
where the $(H_I, H^I)$ denote multi-centered harmonic functions with multiple centers located at $\vec{x}_l$ with electric charges
$q_{l I}$ and magnetic charges $p^I_l$,
\be 
H_I=h_I+\sum_l \frac{q_{l I}}{|\vec{x} - \vec{x}_l|},\quad H^I=h^I + \sum_l \frac{p^I_l}{|\vec{x} - \vec{x}_l|} 
\;,
\label{harmmult}
\ee
with integration constants $h_I \in \mathbb{R}, h^I \in \mathbb{R}$. The quantity $R(\omega)_j$ is expressed in terms of the harmonic functions as
$R(\omega)_j = H^I \partial_j H_I - H_I \partial_j H^I$. Static black holes satisfy $H^I \partial_j H_I - H_I \partial_j H^I=0$ \cite{Behrndt:1997ny}.

These black holes are supported by gauge fields $F_{\mu \nu}^I$ given by \cite{Denef:2000nb,LopesCardoso:2000qm}
\be
F_{tj}^{I-} = - e^g \left[ \partial_j \left({\bar k} X^I \right) + (\partial_j g) \, k {\bar X}^I - \frac{i}{2} e^{2g}\,
R(\omega)_j \left({\bar k} X^I + k {\bar X}^I \right) \right] \;.
\ee
The compensating phase $k$ satisfies the constraint
\be
{\bar k} \partial_j k + i a_j = - \frac{i}{2} e^{2g} \, R(\omega)_j \;,
\label{u1R}
\ee
where $a_{\mu}$ denotes the $U(1)$ connection \eqref{u1a}.

Now consider the linearised solution, which is obtained for large $r = |\vec{x}|$. Denoting
$e^{2g} = 1-  Q/r + O(1/r^2)$, we obtain for the line element, 
\begin{equation}
ds^2 = \eta_{\mu\nu}dx^{\mu}dx^{\nu}+\frac{Q}{r}(dt^2+ dx^i dx^i)  - 2 \omega_i dx^i dt \;.
\label{linelinmc}
\end{equation}
The resulting metric is of 
form $g_{\mu \nu} = \eta_{\mu \nu} + h_{\mu \nu}$, with background metric $\eta_{\mu\nu}=  {\rm diag} (-1, 1, 1, 1)$ and metric fluctuation $h_{\mu \nu}$.
At the linearised level, the harmonic functions $H_I$ and $H^I$ are given by
$H_I = h_I + q_I/r, \, H^I = h^I + p^I/r$, with $q_I = \sum_l q_{lI} , \, p^I = \sum_l p_l^I $, while
$R(\omega)_j$ is given by
\be
R(\omega)_j =  c \, \partial_j \frac{1}{r} \;\;\;,\;\;\; c = h^I q_I - h_I p^I \;.
\label{Romc}
\ee
The gauge fields  $F_{\mu \nu}^I$ become
\be
F_{tj}^{I-} = -  \partial_j \left({\bar k} X^I \right) + \frac{Q}{2} (\partial_j \frac{1}{r}) \, \langle 
k {\bar X}^I  \rangle + \frac{i}{2} 
R(\omega)_j \, \langle {\bar k} X^I + k {\bar X}^I \rangle   \;.
\label{sugraFtiasd}
\ee
The components $F_{tj}^I$ and $F_{ij}^I$ are obtained by linear combinations of $F_{tj}^{I\pm}$. 
Now we recall that
the $U(1)$ connection $a_{\mu}$ vanishes at the linearised level, see \autoref{App:Kahlerconnec}. Then, using \eqref{u1R}, we 
infer ${\bar k} \partial_j k = - \frac{i}{2}  \, R(\omega)_j $ at the linearised level. Using this in \eqref{sugraFtiasd}, we obtain
\be
F_{tj}^{I-} = -  {\bar k} \, \partial_j X^I  + \frac{Q}{2} (\partial_j \frac{1}{r}) \, \langle 
k {\bar X}^I  \rangle + \frac{i}{2} 
R(\omega)_j \, \langle k {\bar X}^I \rangle   \;,
\label{sugraFtiasd2}
\ee
and hence, using \eqref{Romc},
\be
F_{tj}^{I-} = -  {\bar k} \, \partial_j X^I  + \frac{(Q + i c)}{2} (\partial_j \frac{1}{r}) \, \langle 
k {\bar X}^I  \rangle \;.
\label{sugraFtiasd3}
\ee

At the linearised level, the scalar fields $Y^I$ take the form $Y^I = \langle Y^I \rangle + {\cal Y}^I/r$,
where $ \langle Y^I \rangle $ denotes the asymptotic value of $Y^I$, and where  the ${\cal Y}^I$ are expressed in terms
of the charges $(q_I, p^I)$ as well as in terms of the constants $(h_I, h^I)$ that appear in the
harmonic functions \eqref{harmmult}.  Accordingly,
the scalar fields $z^A= Y^A/Y^0$, introduced in \eqref{zIdef}, take the form 
\be
z^A = \langle z^A \rangle + \frac{\mu^A}{r} \;,
\ee
where $\langle z^A \rangle $  denotes the asymptotic value of $z^A$, and the $\mu^A$ are expressed as 
\be
\mu^A =  \frac{{\cal Y}^A -  \langle z^A \rangle \, {\cal Y}^0}{\langle Y^0 \rangle} \;.
\ee
Using \eqref{YX},
the scalar field fluctuations $X^I$ are then given by 
\be
X^I = \frac{\Sigma^I}{r} \;, 
\ee
with
\be
\Sigma^I = {\cal Y }^I \, k - \frac12 Q \langle X^I \rangle \;.
\ee
The values $\Sigma^I$ can thus be determined explicitly in any given model by solving the attractor equations
\eqref{attractor_equations}.

\bibliographystyle{JHEP}

\providecommand{\href}[2]{#2}\begingroup\raggedright\endgroup

\end{document}